\pdfoutput=1
\documentclass[11pt,twoside,a4paper,cmspaper,final,collab]{cms-tdr}
\def\svnVersion{afa841f}\def\svnDate{2024/02/08}\def\cmsCernNoTag{CERN-EP-2023-129}\def\cmsCernDate{\today}\def\cmsMessage{Published in Physical Review Letters as \href{http://dx.doi.org/10.1103/PhysRevLett.132.061801}{\doi{10.1103/PhysRevLett.132.061801}.}}
\begin{document}\cmsNoteHeader{EXO-22-018}

\newlength\cmsTabSkip\setlength{\cmsTabSkip}{1ex}

\newcommand{\mLQ}{\ensuremath{m_\text{LQ}}\xspace}
\newcommand{\mcoll}{\ensuremath{m_\text{coll}}\xspace}
\newcommand{\njets}{\ensuremath{N_\text{jets}}\xspace}
\newcommand{\dphi}{\ensuremath{\abs{\Delta\phi(\ell,\ptvecmiss)}}\xspace}
\newcommand{\Wjets}{\ensuremath{\PW+\text{jets}}\xspace}
\newcommand{\Zjets}{\ensuremath{\PZ+\text{jets}}\xspace}
\newcommand{\taujet}{\ensuremath{\tauh+\text{jet}}\xspace}
\newcommand{\eljet}{\ensuremath{\Pe+\text{jet}}\xspace}
\newcommand{\mujet}{\ensuremath{\Pgm+\text{jet}}\xspace}
\ifthenelse{\boolean{cms@external}}{\providecommand{\cmsLeft}{upper\xspace}}{\providecommand{\cmsLeft}{left\xspace}}
\ifthenelse{\boolean{cms@external}}{\providecommand{\cmsRight}{lower\xspace}}{\providecommand{\cmsRight}{right\xspace}}

\cmsNoteHeader{EXO-22-018}
\title{Search for scalar leptoquarks produced via \texorpdfstring{$\PGt$}{tau}-lepton-quark scattering in \texorpdfstring{$\Pp\Pp$}{pp} collisions at \texorpdfstring{$\sqrt{s}=13\TeV$}{sqrt(s) = 13 TeV}}

\author*[cern]{CMS Collaboration}

\date{\today}

\abstract{
The first search for scalar leptoquarks produced in $\Pgt$-lepton-quark collisions is presented. It is based on a set of proton-proton collision data recorded with the CMS detector at the LHC at a center-of-mass energy of 13\TeV corresponding to an integrated luminosity of 138\fbinv.  The reconstructed final state consists of a jet, significant missing transverse momentum, and a \PGt lepton reconstructed through its hadronic or leptonic decays. Limits are set on the product of the leptoquark production cross section and branching fraction and interpreted as exclusions in the plane of the leptoquark mass and the leptoquark-$\PGt$-quark coupling strength. 
}

\hypersetup{
pdfauthor={CMS Collaboration},
pdftitle={Search for scalar leptoquarks produced via tau-lepton-quark scattering in pp collisions at sqrt(s) = 13 TeV},
pdfsubject={CMS},
pdfkeywords={CMS, leptoquarks, taus}}

\maketitle 

Leptoquarks (LQs) are hypothetical color-triplet bosons carrying both baryon and lepton numbers and having fractional electric 
charge~\cite{PhysRevD.10.275,PhysRevLett.32.438,DIMOPOULOS1979237,DIMOPOULOS198069,SCHREMPP1985101,BUCHMULLER1986377}.
Mechanisms involving LQs coupled to third-generation quarks and leptons 
could explain the deviations from standard model (SM) predictions observed in several measurements
of B meson decays, as explained in Refs.~\cite{Dumont:2016xpj,Buttazzo:2017ixm,Kumar:2018kmr,Angelescu:2021lln} and references therein.

At the CERN LHC, LQs can be produced singly through quark-gluon fusion, in pairs through gluon fusion or quark-antiquark annihilation,
as part of the nonresonant production of two leptons in the $t$-channel, or from lepton-quark collisions.
While the ATLAS and CMS Collaborations have performed searches for LQs targeting the first three production
mechanisms~\cite{ATLAS:2021jyv,CMS:2018txo,CMS:2020wzx,ATLAS:2021oiz,CMS:2022goy,CMS:2022ncp,ATLAS:2023kek},
the last production mode has never been explored at the LHC.
Recent theoretical progress in the determination of the lepton and photon density functions in the proton~\cite{Buonocore:2020nai},
based on the LUX approach~\cite{Manohar:2016nzj,Manohar:2017eqh}, 
has shown that a significant LQ production cross section can be expected in proton-proton ($\Pp\Pp$) 
collisions~\cite{Buonocore:2020erb,Haisch:2020xjd,Greljo:2020tgv,Buonocore:2022msy}.

\begin{figure}[ht]
  \centering
  \includegraphics[width=0.3\textwidth]{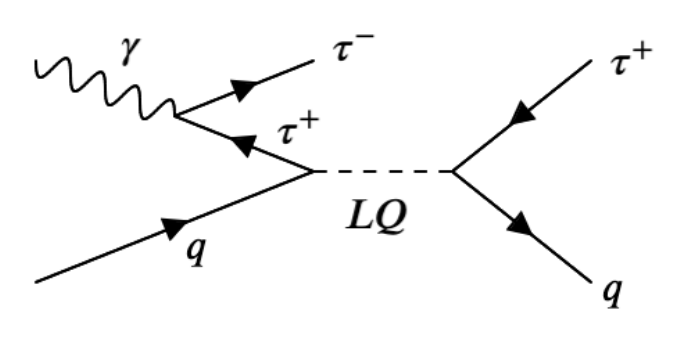}
  \caption{\label{fig:feynman} Feynman diagram of the lepton-induced LQ production.}
\end{figure}

Lepton-induced production of LQs proceeds via the collision of a lepton and a quark, where the lepton is 
produced through a photon via quantum fluctuations in the proton. A leading-order (LO) Feynman diagram for this process is shown in 
Fig.~\ref{fig:feynman}. After the decay of the
LQ, the final state consists of a high transverse momentum (\pt) centrally-produced lepton from the LQ decay,
a same-flavor opposite-sign forward lepton from the photon decay, and a high-\pt centrally-produced jet from the LQ decay. 
The forward lepton in this production mode usually has $\pt$ below reconstruction threshold and differs from other 
production modes, which typically have two high-\pt leptons. 
This Letter presents the first search for scalar LQs produced in $\Pgt$q collisions. 
For the strongest experimental sensitivity, we consider couplings 
to light-flavor ($\PQu$, $\PQd$, and $\PQs$) and \PQb quarks. 
The search is based on $\Pp\Pp$ collision data at $\sqrt{s}=13\TeV$ collected with the CMS detector in 2016--2018 
corresponding to an integrated luminosity of 138\fbinv. 
Three different decay modes of the \PGt lepton are studied, resulting in the $\taujet$, $\eljet$, and $\mujet$ final states, where $\tauh$ denotes a
\PGt lepton decaying hadronically.
Tabulated results are provided in the HEPData record for this analysis~\cite{hepdata}.

The CMS apparatus~\cite{CMS:2008xjf} is a multipurpose, nearly hermetic detector, designed to trigger on~\cite{CMS:2020cmk,CMS:2016ngn} and identify electrons, muons, photons, and both charged and neutral hadrons~\cite{CMS:2020uim,CMS:2018rym,CMS:2014pgm}. A global ``particle-flow" (PF) algorithm~\cite{CMS:2017yfk} aims to reconstruct all individual particles in an event, combining information provided by the all-silicon inner tracker and by the crystal electromagnetic, and a brass and scintillator hadron calorimeters, operating inside a 3.8\unit{T} superconducting solenoid, with data from the gas-ionization muon detectors embedded in the flux-return yoke outside the solenoid. The reconstructed particles are used to build \PGt leptons, jets, and missing transverse momentum ($\ptvecmiss$)~\cite{CMS:2018jrd,CMS:2016lmd,CMS:2019ctu,CMS:2022prd}.

The signal and some of the background contributions are estimated using simulations. 
Signal samples for scalar LQs in the $s$-channel single LQ production are
generated at next-to-LO (NLO) with \POWHEG interfaced with \HERWIG~\cite{Bahr:2008pv} for the showering, using the models
 and \textsc{LUXlep} lepton PDFs described in Refs.~\cite{Buonocore:2020erb,Buonocore:2020nai}. The
\textsc{LUXlep} PDF set results from the combination of the
\textsc{NNPDF3.1luxQED} set~\cite{Bertone:2017bme} and the lepton PDFs of Ref.~\cite{Buonocore:2020nai},
which are obtained with a framework similar to that used to determine the PDFs of the photon in the proton using
electron-proton scattering data~\cite{Manohar:2016nzj,Manohar:2017eqh}.
The signal samples are generated for $\PQb\PGt$ ($\PQu\PGt$) couplings, for LQ masses $\mLQ$ between 0.6 and 2.0 (0.6 and 3.0)\TeV,
an absolute electric charge of $2/3e$ ($1/3e$),
and a Yukawa coupling at the LQ-lepton-quark vertex, $\lambda$, between 0.5 and 3.0 (0.2 and 2.0). Charge conjugate pairs are simulated
together.
The analysis acceptance and efficiency, determined from simulation, are similar for couplings to any light-flavor quark,
and the samples with
LQ-$\PQu$-$\PGt$ couplings are also used to extract results on LQ-$\PQd$-$\PGt$ and LQ-$\PQs$-$\PGt$ couplings.
Results on LQ-$\PQc$-$\PGt$ couplings cannot be extracted from the same simulation because events with such 
couplings have a different efficiency for the requirement on the identification of jets originating from {\PQb} quarks, 
which is used to build analysis categories as described later. 

The cross sections for all coupling hypotheses are computed at NLO~\cite{Greljo:2020tgv,Buonocore:2022msy}.
The \MGvATNLO 2.6.5 event generator~\cite{Alwall:2014hca} is used to generate events originating from $\Zjets$ and $\Wjets$ 
processes. They are simulated
 at NLO with the FxFx jet matching and merging~\cite{Frederix:2012ps}. 
The vector boson \pt distribution is corrected to match a calculation at next-to-NLO (NNLO) for strong interactions and NLO with the 
next-to-leading logarithmic Sudakov approximation for electroweak interactions~\cite{Lindert:2017olm}. 
The \MGvATNLO generator is also used for the simulation of diboson production, while 
\POWHEG 2.0~\cite{Nason:2004rx,Frixione:2007vw, Alioli:2010xd, Alioli:2010xa, Alioli:2008tz} is
 used for \ttbar and single top quark production. The generators are interfaced with
 \PYTHIA 8.240~\cite{Sjostrand:2014zea} to model the parton showering, fragmentation, and hadronization, as well as the decay of the
 \PGt leptons. The \PYTHIA parameters affecting the description of the underlying event are set to
 the {CP5} tune~\cite{CMS:2019csb}.
The \textsc{NNPDF3.1} parton distribution function (PDF) set~\cite{Ball:2011uy,Ball:2013hta,Ball:2017nwa} at NNLO precision 
is used for background simulations.
Additional $\Pp\Pp$ interactions per bunch crossing are added to the simulated samples with the frequency distribution 
matching that observed in data.
Generated events are processed through a \GEANTfour~\cite{Agostinelli:2002hh} simulation of the CMS detector.

Electrons are reconstructed from energy deposits in the calorimeters and tracks in the tracking system,
and identified with a cut-based discriminant~\cite{CMS:2020uim}.
Muons are reconstructed from tracks and hits in the tracker and muon systems~\cite{CMS:2018rym,CMS:2019ied}.
Jets are clustered from PF candidates using the anti-$\kt$ \FASTJET algorithm with a distance parameter $R$ of 0.4~\cite{Cacciari:2008gp,CMS-PAS-JME-16-003}. Their energy is corrected on an event-by-event
basis~\cite{CMS-DP-2020-019}.
Jets originating from {\PQb} quarks are identified with the medium working point of the \textsc{DeepJet} algorithm~\cite{Sirunyan:2017ezt,Bols:2020bkb}.
The decay products of \PGt leptons decaying hadronically are reconstructed with the hadrons-plus-strips algorithm~\cite{CMS:2018jrd}. 
Quark and gluon jets, electrons, and muons misidentified as $\tauh$ candidates are reduced with deep neural network discriminants~\cite{CMS:2022prd}. 
The tight working point is used to separate $\tauh$ candidates from jets; its efficiency is about 
75\% for $\tauh$ with $\pt>100\GeV$~\cite{CMS:2022prd}. The loosest working point, used in the background estimation procedure, has an efficiency 
above 98\%. 
The vector $\ptvecmiss$ is defined as the projection onto the plane perpendicular to the beam axis of the negative vector momentum sum of all reconstructed PF objects in an event. Its magnitude is referred to as $\ptmiss$.

The selection of events in the $\taujet$ final state relies either on a single $\tauh$ trigger with online thresholds
ranging from 140 to 180\GeV, depending on the data-taking year, or on a trigger requiring
the scalar \pt sum of jets in the event with $\pt > 40\GeV$ to be above 900--1050\GeV.
Events are selected in the $\eljet$ final state using a single isolated electron 
trigger with online thresholds in the range of 27--35\GeV complemented with a photon trigger for $\pt>175-200\GeV$.
In the $\mujet$ final state, single muon triggers with online thresholds between 24 and 50\GeV are used. 
Depending on whether the jet is tagged as originating from a \PQb quark~\cite{Sirunyan:2017ezt,Bols:2020bkb}, 
the event is classified to be in a ``btag" or ``no-btag" category. 
A significant $\ptvecmiss$ satisfying $\dphi<0.2$--0.3 is required, where $\ell$ stands for the visible decay products of the \PGt lepton, 
as expected from the presence of one or more neutrinos in the \PGt lepton decay.
The offline selection criteria for the lepton, leading jet, and $\ptvecmiss$ are presented in Table~\ref{tab:sel}. 
They differ slightly for final states with leptonically and hadronically decaying \PGt leptons 
because of the different background contributions, triggers, and fractions of visible \PGt lepton momentum. 
To select the LQ production mechanism targeted in this analysis, events are discarded if a second well-identified and isolated electron, 
muon, or $\tauh$ candidate with $\pt>50\GeV$ and absolute pseudorapidity, $\abs{\eta}$, less than 2.1 is found.
This additional lepton veto removes about 5\% of the signal events in 
simulation. With this veto, other LQ production modes contribute less than 1\% to the signal yield and are neglected.

\begin{table}
\centering
\topcaption{Selection criteria per final state. }\label{tab:sel}
\renewcommand{\arraystretch}{1.2}
\begin{scotch}{lccc}
\multicolumn{1}{c}{Variable} & $\taujet$ & $\eljet$ & $\mujet$ \\
\hline
$\pt^\ell (\GeVns)$ & $>$200 & $>$100 & $>$100 \\
$\abs{\eta^\ell}$ & $<$2.1 & $<$2.1 & $<$2.1 \\
$\pt^\text{jet} (\GeVns)$ & $>$300 & $>$200 & $>$200 \\
$\abs{\eta^\text{jet}}$ & $<$2.4 & $<$2.4 & $<$2.4 \\
$\ptmiss (\GeVns)$ & $>$100 & $>$150 & $>$150 \\
$\pt(\vec{\ell}+\ptvecmiss) (\GeVns)$ & $>$100 & $>$100 & $>$100 \\
$\dphi$ (radians) & $<$0.3 & $<$0.2 & $<$0.2 \\
$\Delta R(\ell,\text{jet})$ & $>$0.5 & $>$0.5 & $>$0.5 \\
\end{scotch}
\end{table}

As an estimation of $\mLQ$ at reconstruction level, we calculate the collinear mass, defined as $\mcoll=m_\text{vis}(\PGt, \text{jet}) / \sqrt{x_\text{vis}}$,
where $m_\text{vis}(\PGt, \text{jet})$ is the invariant mass of the visible \PGt decay products and the jet,
$x_\text{vis}=\pt^\text{vis}(\PGt) / (\pt^\text{vis}(\PGt) + \pt^\text{invis}(\PGt))$, and
$\pt^\text{invis}(\PGt)$ is the component of $\ptvecmiss$ in the direction of the visible \PGt decay products.
This calculation assumes that the neutrinos from \PGt lepton decays are the only source of $\ptvecmiss$ and have the same $\eta$ as the
visible \PGt decay products. 
The experimental resolution on $\mcoll$ is about 5--10\%, depending on the final state and $\mLQ$, and 
for $\lambda\lesssim1$ the natural width of the signal is negligible by comparison. In simulated events, the $\mcoll$ distribution
peaks at the generated $\mLQ$ values.

Background events with a prompt lepton in the final state ($\Wjets$, Drell--Yan, diboson, single top quark, and $\ttbar$
production events) are estimated from simulation and normalized to their theoretical cross 
sections~\cite{Gavin:2012sy,Gavin:2010az,Czakon:2011xx,Gehrmann:2014fva,Campbell:2020fhf,Kidonakis:2021vob}, with the exception of  
the normalization of $\Wjets$ events in the btag categories. Because of the lack of a precise
prediction for the cross section of $\Wjets$ in association with
heavy-flavor jets, the latter is extracted from 
data in control regions (CRs) in the $\eljet$ and $\mujet$ final states with $0.2<\dphi<0.4$, 
where $\phi$ is the azimuthal angle in radians.
The variable $\njets$, defined as the number of jets with 
$\pt>30\GeV$ and $\abs{\eta}<2.4$, and separated 
by $\Delta R = \sqrt{\smash[b]{(\Delta \eta)^2 + (\Delta \phi)^2}} > 0.5$ from the selected lepton, 
is used as an observable to separate $\Wjets$ from \ttbar events, which on average have a higher number of jets. After the 
maximum likelihood fit including the signal regions (SRs), described later, 
the cross section for the $\Wjets$ background in the btag categories is measured to be $1.04^{+0.07}_{-0.06}$ 
times the prediction from simulation, using the inclusive $\Wjets$ cross section at NNLO~\cite{Gavin:2012sy}.

The backgrounds with a jet misidentified as a $\tauh$ candidate (an electron or muon), dominated 
by quantum chromodynamics (QCD) multijet events, constitute about 50 (10)\% of the total background and are estimated from data. 
The probability for jets selected with the loosest working point of the  $\tauh$ discriminator
to pass the tighter identification requirements in 
the SR, called the misidentification factor (MF), is measured as a function of $\pt(\tauh)$, $\njets$, $\abs{\eta}$, and jet \pt. 
It is in the range 0.10--0.25. This measurement is performed in a CR 
with a selection identical to that in the SR, except 
that the $\ptmiss$ requirement is inverted. A multiplicative correction for the extrapolation to higher $\ptmiss$ values in the SR 
is applied.  
The MF for electrons is measured as a function of the electron \pt because the trigger requirements are \pt-dependent, 
while the muon MF is measured as a function of $\ptmiss$, since the relative fraction of QCD multijet and $\Wjets$ backgrounds varies 
with $\ptmiss$.  
The observed $\ptmiss$ linear dependence is extrapolated to the SR for $150<\ptmiss<300\GeV$ and is assumed constant above 
300\GeV. 
Events that 
pass the SR selection with the exception of the $\tauh$, electron, or muon identification, depending on the final state, 
are reweighted using the MFs to estimate the background with misidentified jets. Contributions from prompt $\tauh$ candidates, electrons, or muons in this region are 
estimated from simulation and subtracted to avoid double-counting events containing genuine leptons. 

After the application of the selection criteria shown in Table~\ref{tab:sel}, boosted decision trees (BDTs) are trained with the \textsc{tmva} package~\cite{2007physics...3039H} for each
final state to improve the separation between the signal and background. The input
variables, chosen to have a limited dependence on $\mLQ$, are the following: $\abs{\Delta\phi}$ between pairs of
analysis objects (electrons, muons, $\tauh$, jets, $\ptvecmiss$), the ratio of the objects' \pt to $\mcoll$ or to each other's \pt,
 the $\Delta R$ separation between the jet and the reconstructed \PGt lepton, and $\njets$.
The BDTs are trained with a mixture of all signal samples with various $\mLQ$ and $\lambda$, against background
events coming from $\Wjets$ production and either processes with misidentified jets in the $\taujet$ channel, or
$\ttbar$ events in the $\eljet$ and $\mujet$ channels. The BDT training is performed once for each final state, and is used in the btag and no-btag categories, 
for all data-taking years and $\mLQ$ hypotheses. The BDT output distribution is verified to be well
predicted in a validation region where the requirement on $\dphi$ is inverted.

Events in the no-btag\,(btag) category are further split into 4\,(3) subcategories on the basis of their BDT output. In the btag 
subcategories, events with $\njets>2$ are
vetoed to reduce the contribution from the \ttbar background. 
The discriminating observable is $\mcoll$ in all subcategories and final states.

Uncertainties in the reconstruction, identification, and isolation of $\tauh$ candidates (electrons, muons) 
are determined via the ``tag-and-probe" method~\cite{Khachatryan:2010xn} and can be as high as 15\,(2, 2)\%. The \PQb tagging 
uncertainties for heavy-flavor jets and mistagging uncertainties for light-flavor quark and gluon jets are included 
with partial correlations between the data-taking years. They are in the range 6--9\% in the btag category for 
LQs coupled to \PQb quarks. The uncertainty in the efficiency of the 
$\tauh$ (electron, muon) trigger ranges up to 11\,(2, 2)\%. 
The uncertainty in the energy scale of $\tauh$ candidates is below the percent level for all $\tauh$ decay modes, and is propagated to the 
$\mcoll$ distributions. The impact of uncertainties in the energy scale and resolution of jets, and in the measurement of $\ptvecmiss$, is smaller 
by comparison. 
The uncertainties in the energy scale of electrons and muons are negligible as compared with the uncertainties mentioned above. 

The uncertainties in the \ttbar, diboson, and single top quark production cross sections are 5.2, 2.5, and 3.7\%, respectively~\cite{Czakon:2011xx,Gehrmann:2014fva,Campbell:2020fhf,Kidonakis:2021vob}. The
uncertainties in the NNLO cross sections of the $\Zjets$ and $\Wjets$ backgrounds are 2 and 3\%, respectively~\cite{Gavin:2012sy,Gavin:2010az}. In the btag category, the $\Zjets$ background normalization uncertainty is increased to 20\% on the basis of the agreement observed in 
a CR with 2 $\tauh$ candidates and 1 \PQb tagged jet, while the normalization of the 
$\Wjets$ background is left floating in the fit, resulting in an uncertainty of 7\% as detailed above.
The acceptance uncertainties in the renormalization and factorization scales, PDFs, and parton showering 
for these simulated processes are also included.
For the signal, uncertainties in the renormalization and factorization scales, and in the PDFs are included. They affect the $\mcoll$ 
distributions and have a normalization effect in the ranges 3--4\,(4--6)\% (scales) and 2--3\,(2--4)\% (PDFs), respectively, for 
LQs with $\PQu\PGt$ ($\PQb\PGt$) couplings, in the mass ranges considered in the analysis~\cite{Greljo:2020tgv}.
The uncertainty in the integrated luminosity is 1.6\%~\cite{CMS-LUM-17-003,CMS-PAS-LUM-17-004,CMS-PAS-LUM-18-002}.

Different uncertainties affecting the $\mcoll$ distributions are considered for the background with jets misidentified as 
$\tauh$ candidates: 10\% uncertainty per detector region (barrel or endcaps), 10\% uncertainty per $\njets$ bin, 
30\% uncertainty for events with $\pt(\tauh) > 600\GeV$, 10\% uncertainty in the leading jet \pt correction, 
and the $\ptmiss$ correction uncertainty described earlier.  
The uncertainty in the MFs for electrons and muons originates from the limited number of events in the measurement and 
from the choice of observables and selection criteria. 
It results in a 20\% normalization uncertainty for backgrounds with misidentified electrons or muons, uncorrelated between the data-taking years. 
In the $\mujet$ channel, an uncertainty in the $\ptmiss$ dependence is included. 

\begin{figure*}[htb!]
  \centering
  \includegraphics[width=0.3\textwidth]{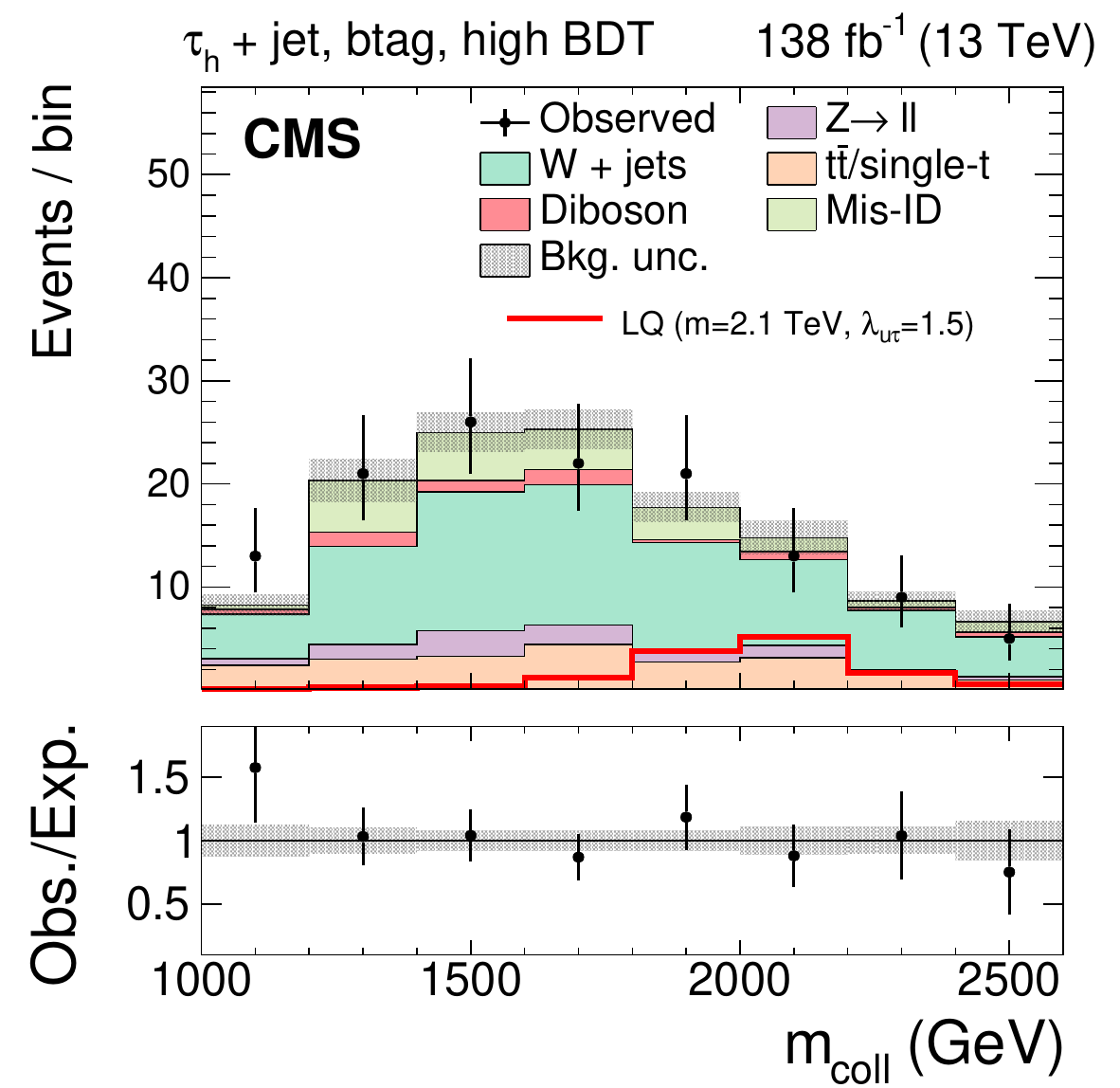}
  \includegraphics[width=0.3\textwidth]{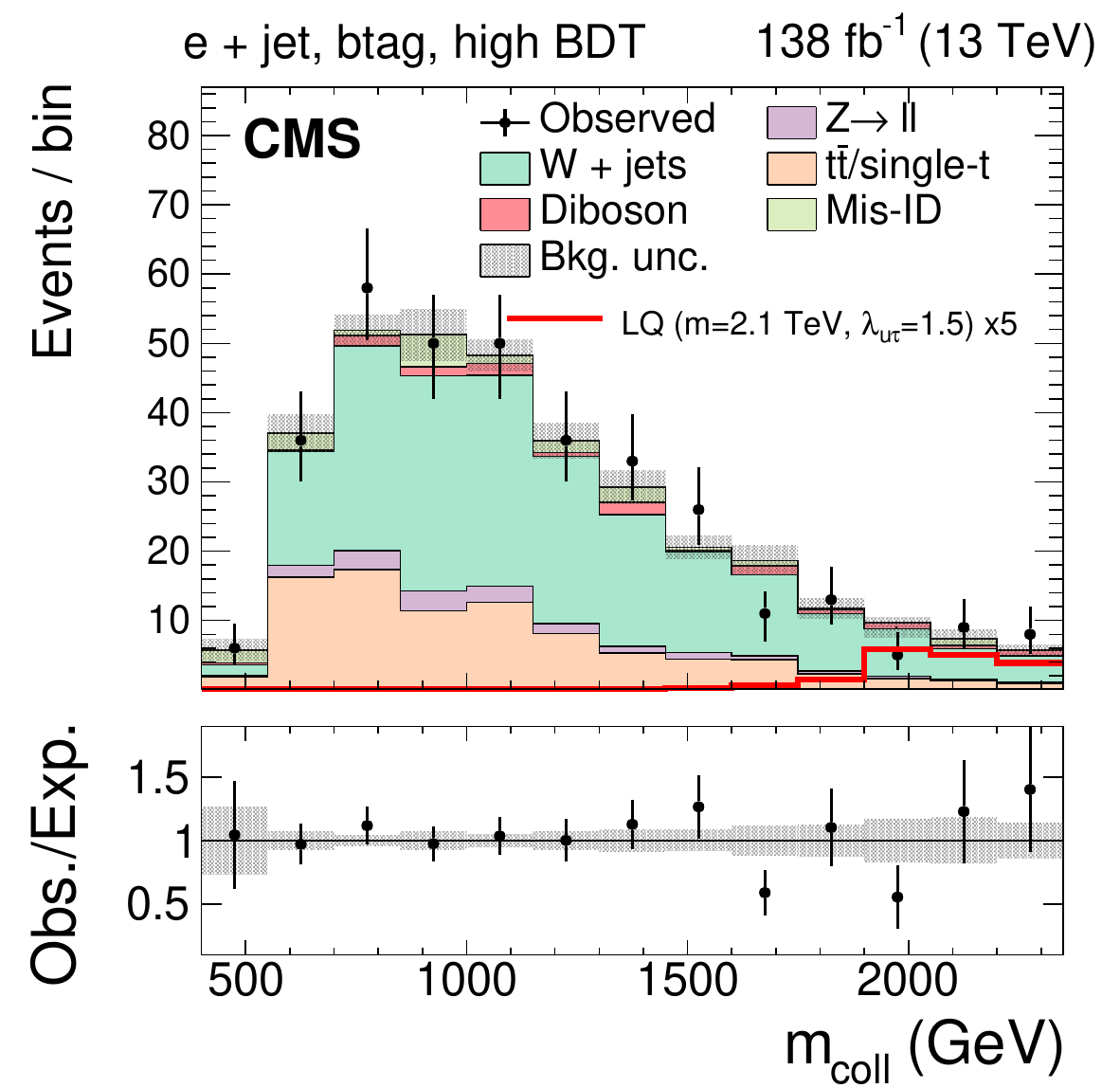}
  \includegraphics[width=0.3\textwidth]{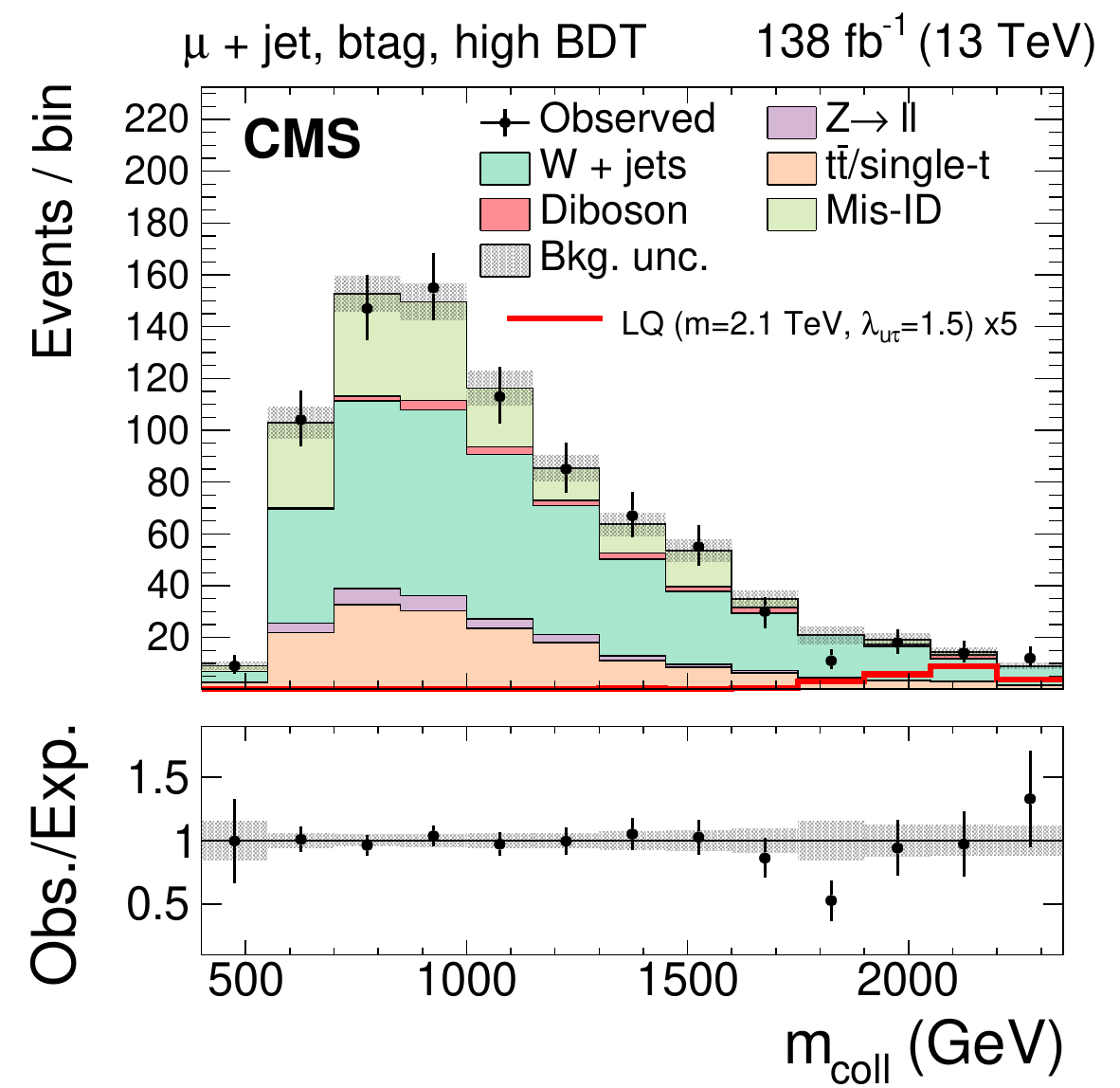}\\
  \includegraphics[width=0.3\textwidth]{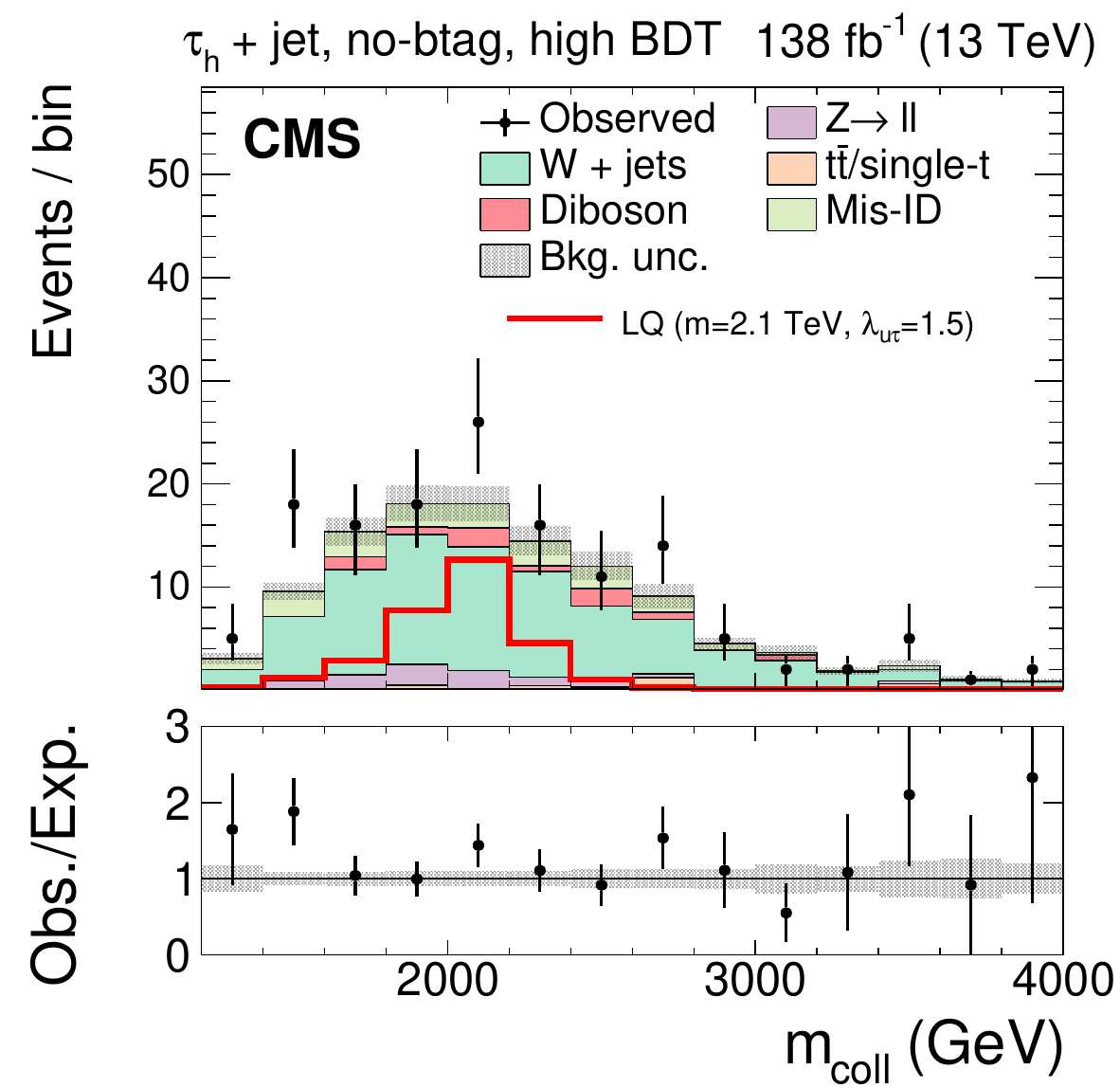}
  \includegraphics[width=0.3\textwidth]{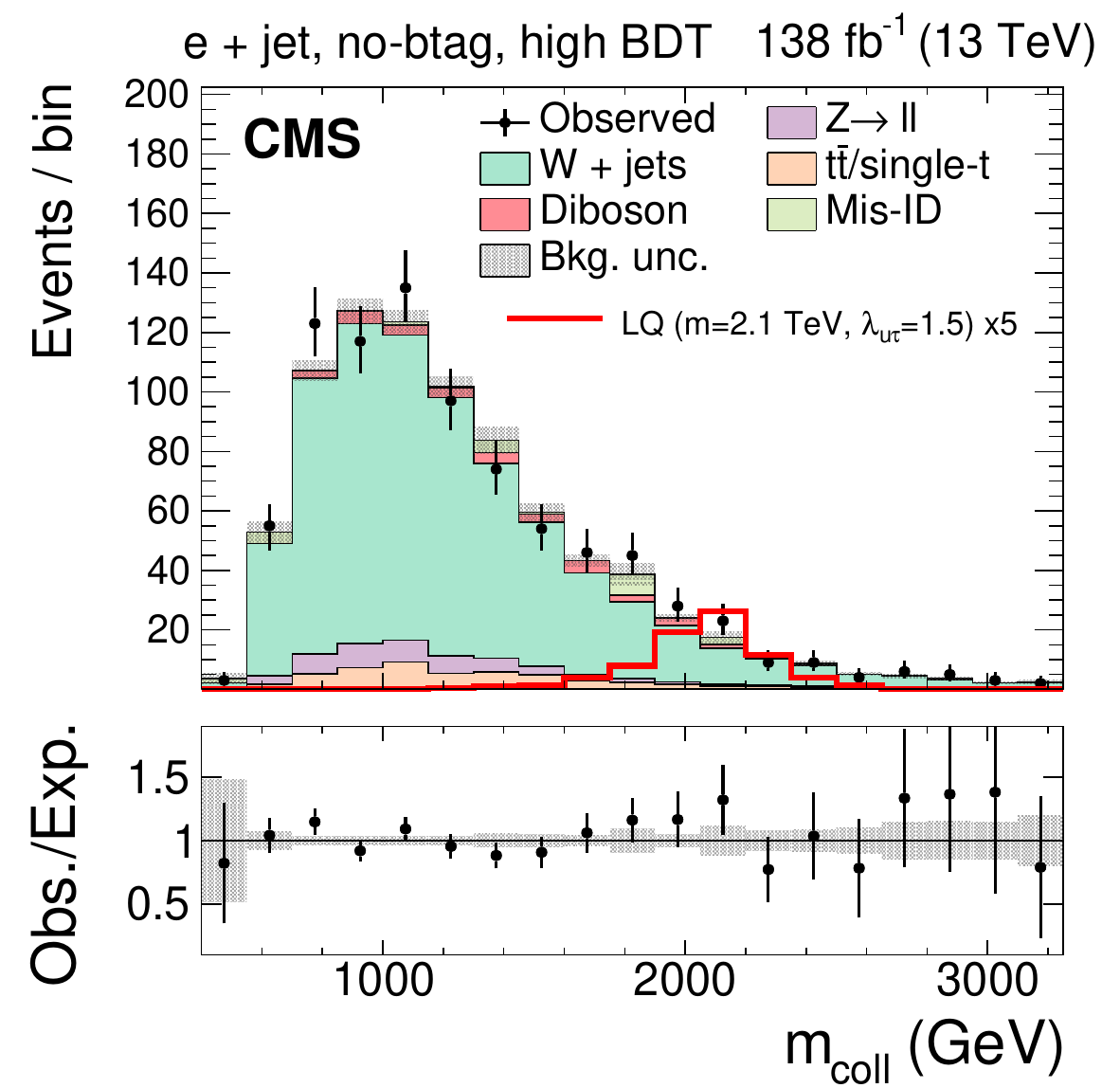}
  \includegraphics[width=0.3\textwidth]{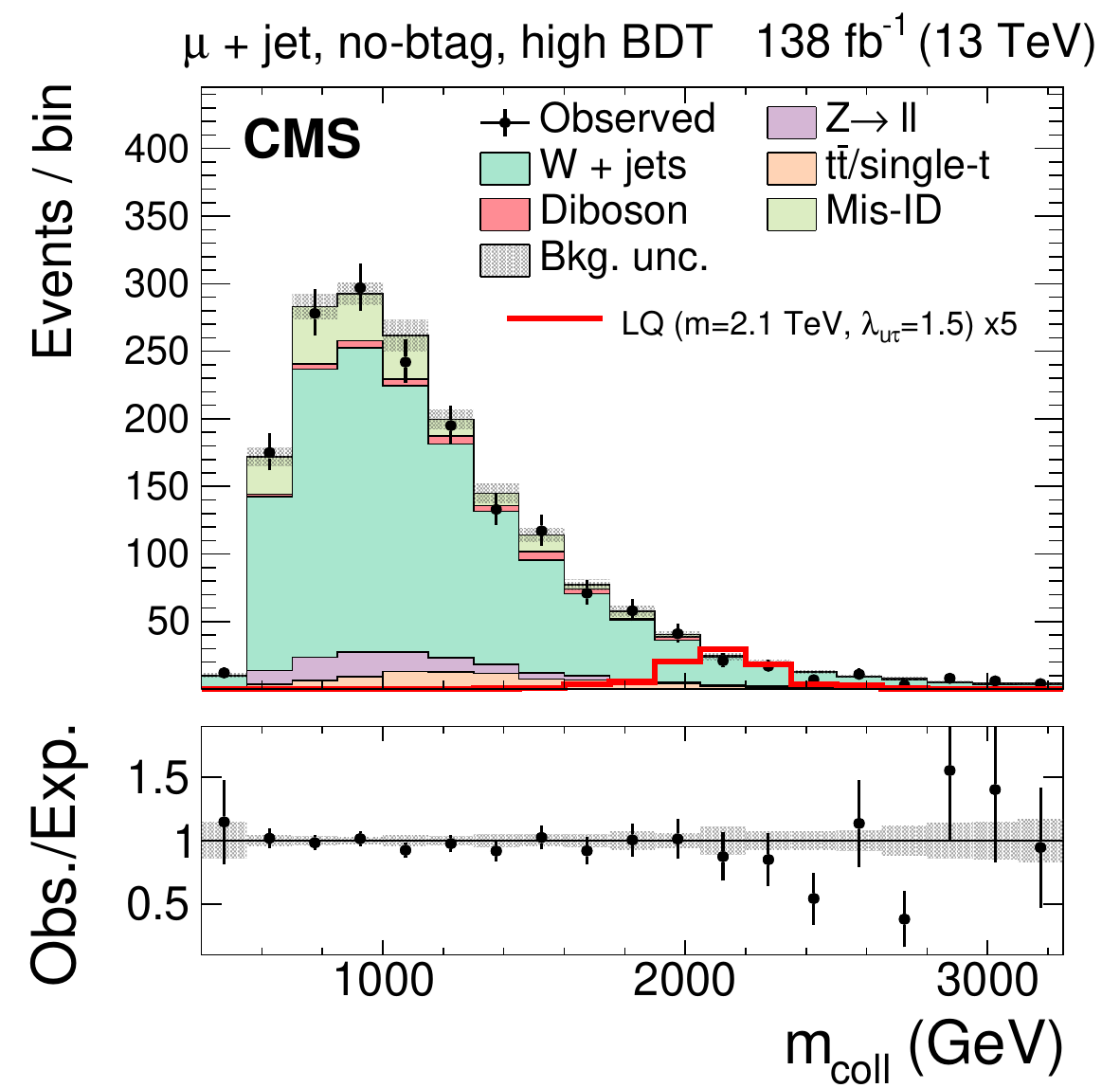}\\
  \caption{\label{fig:mcoll} Observed and expected distributions of $\mcoll$ in the $\taujet$ (left), $\eljet$ (center), and $\mujet$ (right) channels for the btag (upper) and no-btag (lower) subcategories with the BDT requirements selecting the most signal-like events. The bands include statistical and systematic uncertainties. The background distributions are the results of the maximum likelihood fit. For clarity, the signal distributions in the $\eljet$ and $\mujet$ final states are multiplied by a factor of 5.}
\end{figure*}

A maximum likelihood fit is performed with the signal normalization as a free parameter, and the systematic uncertainties
described above as nuisance parameters. The $\mcoll$ distributions in the different subcategories and final states are fitted 
simultaneously, together with the $\njets$ distributions in the CRs that control the normalization 
of the background from $\Wjets$ with a \PQb tagged jet. Those with the highest BDT output requirement, selecting the most 
signal-like events, are shown in Fig.~\ref{fig:mcoll}. 
The btag subcategory with the highest BDT output requirement contains about
70 (41, 56)\% of the LQ events with $\mLQ=2\TeV$ and $\lambda_{\PQb\PGt}=1.0$ entering the btag category, 
and about 9.8 (5.6, 12)\% of the total background in the $\taujet$ ($\eljet$, $\mujet$) final state.
The no-btag subcategory with the highest BDT output requirement contains about 
55 (30, 33)\% of the LQ events with $\mLQ=3\TeV$ and $\lambda_{\PQu\PGt}=1.0$ entering the no-btag category, 
and about 0.86 (1.1, 1.9)\% of the total background 
in the no-btag category in the $\taujet$ ($\eljet$, $\mujet$) final state.

\begin{figure}[htb!]
  \centering
  \includegraphics[width=0.49\textwidth]{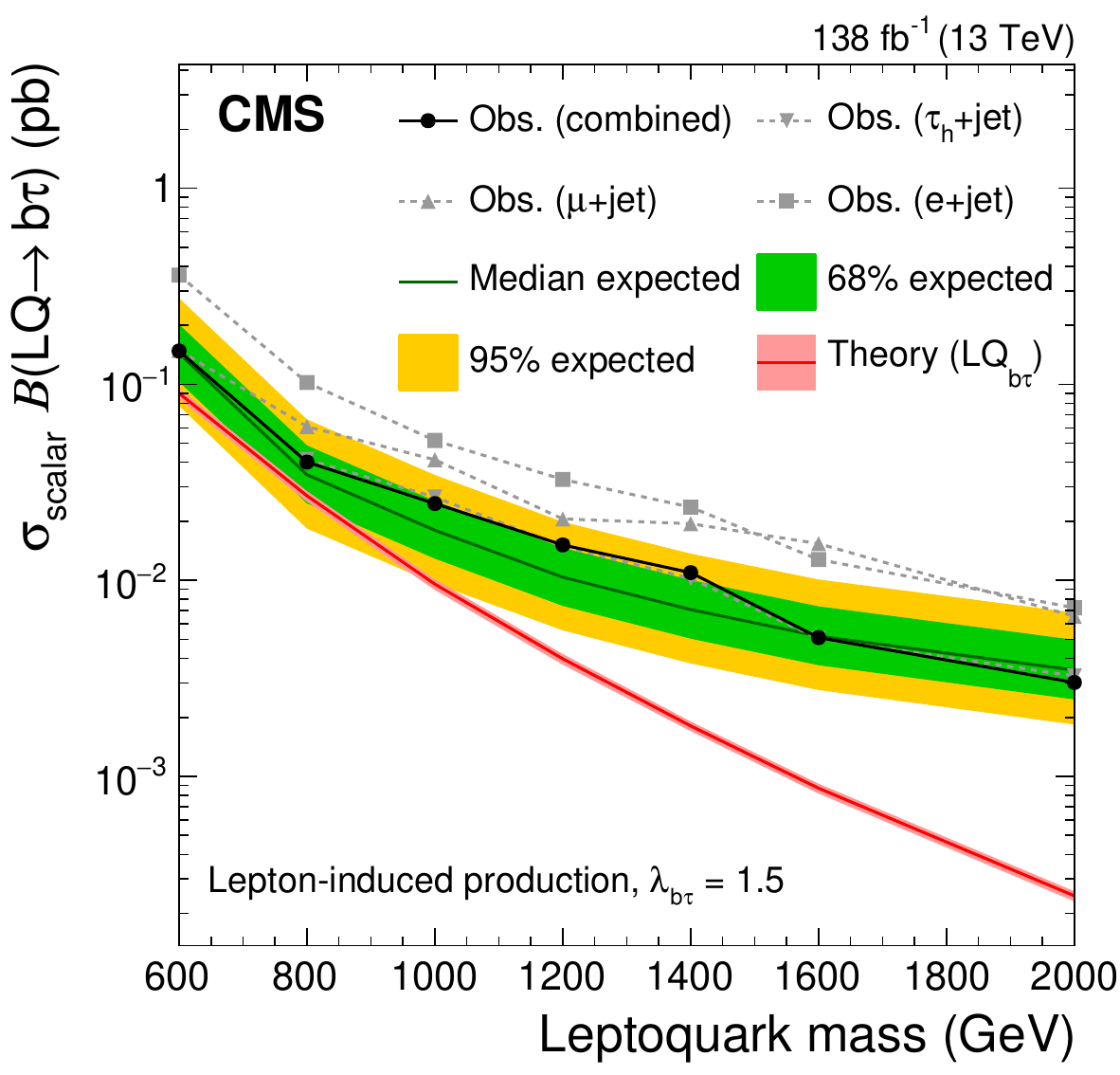}
  \includegraphics[width=0.49\textwidth]{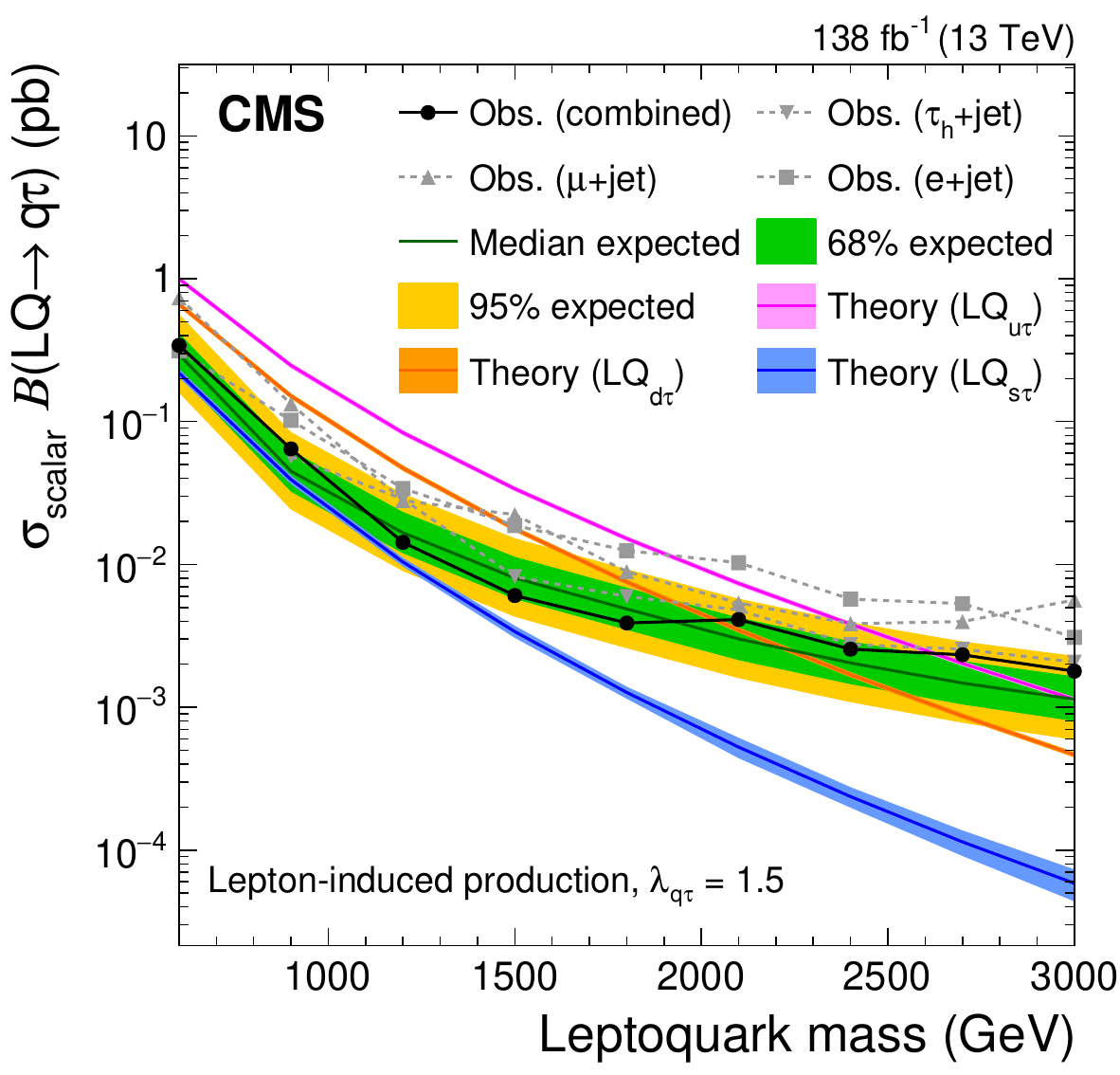}
  \caption{\label{fig:limits} Expected and observed upper limits at 95\% \CL on the product of the scalar lepton-induced LQ production cross section and the branching fraction for a LQ coupled to \PQb quarks and \PGt leptons (\cmsLeft), or to light-flavor quarks and \PGt leptons (\cmsRight), using $\lambda=1.5$. The theoretical cross sections correspond to the calculations of Refs.~\cite{Greljo:2020tgv,Buonocore:2022msy}. The inner (green) band and the outer (yellow) band indicate the regions containing 68 and 95\%, respectively, of the distribution of limits expected under the background-only hypothesis. The filled circles show the observed limits for the combination of final states, while the other markers indicate the observed results per final state.}
\end{figure}

\begin{figure}[htb!]
  \centering
  \includegraphics[width=0.49\textwidth]{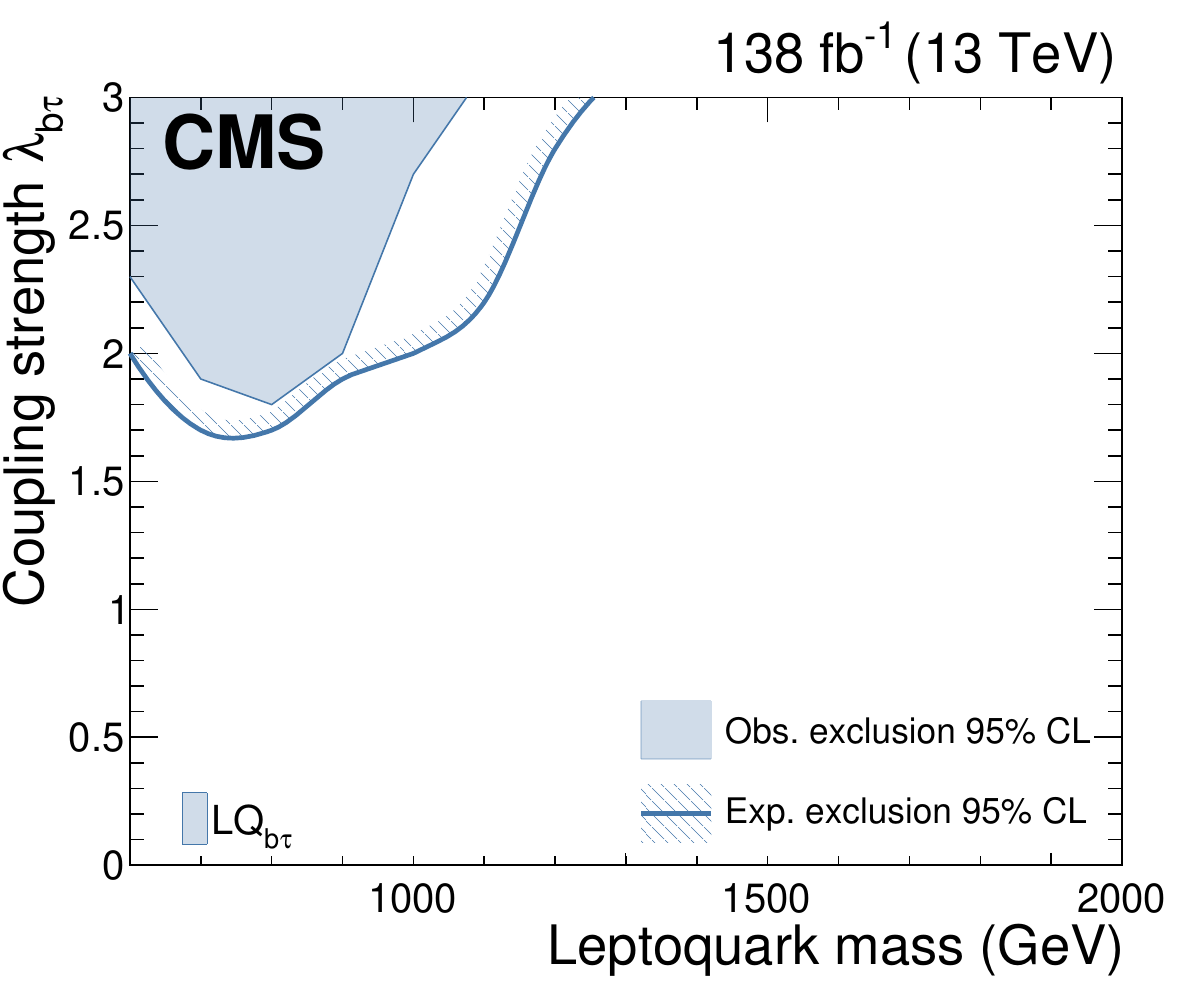}
  \includegraphics[width=0.49\textwidth]{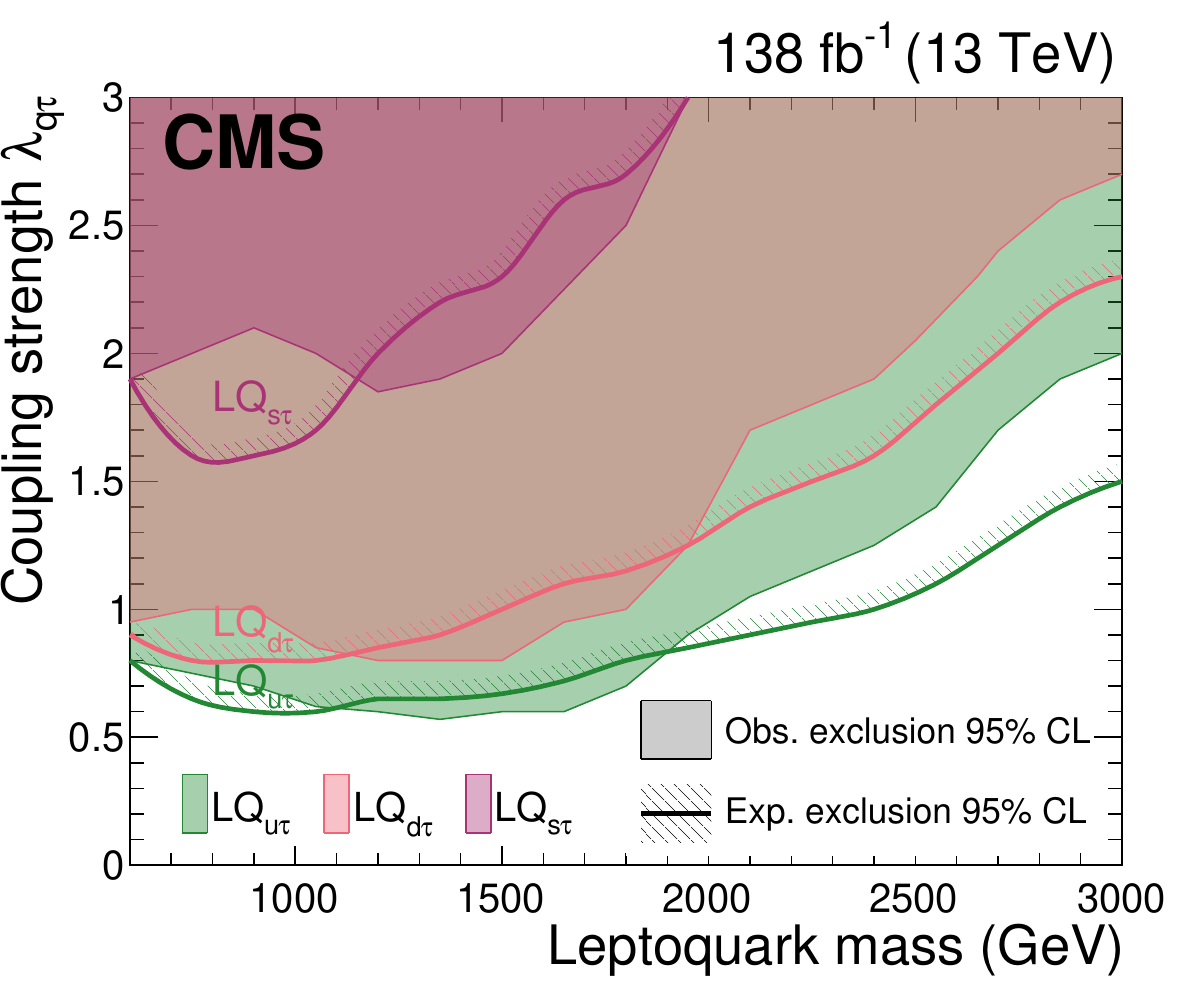}
  \caption{\label{fig:modeldependent} Upper limit at 95\% \CL on the coupling strength $\lambda$ of a scalar LQ to \PQb quarks and \PGt leptons (\cmsLeft), and to light-flavor quarks and \PGt leptons (\cmsRight). Regions above the hatched lines correspond to the expected exclusions.}
\end{figure}

No statistically significant excess above the SM backgrounds
is observed.  
Upper limits at 95\% confidence level (\CL) are set on the product of the LQ production cross section and the branching fraction 
for different 
coupling hypotheses, using the \CLs method~\cite{Junk:1999kv,Read:2002hq} in the asymptotic approximation~\cite{Cowan:2010js}. 
Because of the trigger requirements, the signal acceptance in the $\taujet$ final state is low for $\mLQ=600\GeV$ and the limits for this mass point are derived from the distributions in the $\eljet$ and $\mujet$ final states only. 
The 95\% \CL limits are in the range 0.34--0.0018 (0.15--0.0030)\unit{pb}  
for masses between 0.6 and 3.0 (0.6 and 2.0)\TeV for $\lambda_{\PQu\PGt}=1.5$ ($\lambda_{\PQb\PGt}=1.5$), as shown in 
Fig.~\ref{fig:limits}. 
The limits are translated into exclusion regions in the $\mLQ$-$\lambda$ plane, as shown in Fig.~\ref{fig:modeldependent}, 
assuming the branching fraction of the LQs to a quark and a \PGt lepton to be 100\% for the considered quark flavor. 
The observed limits on LQs coupling to \PQb quarks and \PGt leptons extend existing constraints from searches in other production 
modes at high $\mLQ$~\cite{CMS:2018txo}.  
Leptoquarks coupling to light-flavor quarks and \PGt leptons can be excluded at masses 
above the previously existing limit of 1.3\TeV~\cite{ATLAS:2023kek}, for $\lambda_{\PQu\PGt}$ ($\lambda_{\PQd\PGt}$, $\lambda_{\PQs\PGt}$) 
above 0.6 (0.8, 1.9).

In summary, a search for leptoquarks produced in lepton-quark collisions and coupled to \PGt leptons has
been performed for the first time, using data collected with the CMS detector in 2016--2018.
These limits are complementary to those set using other production
modes at high mass and coupling values for $\PQb\PGt$ couplings, while the limits on the couplings of 
leptoquarks to light-flavor quarks extend the mass range excluded by previous searches in other production modes.

\begin{acknowledgments}
  We congratulate our colleagues in the CERN accelerator departments for the excellent performance of the LHC and thank the technical and administrative staffs at CERN and at other CMS institutes for their contributions to the success of the CMS effort. In addition, we gratefully acknowledge the computing centers and personnel of the Worldwide LHC Computing Grid and other centers for delivering so effectively the computing infrastructure essential to our analyses. Finally, we acknowledge the enduring support for the construction and operation of the LHC, the CMS detector, and the supporting computing infrastructure provided by the following funding agencies: SC (Armenia), BMBWF and FWF (Austria); FNRS and FWO (Belgium); CNPq, CAPES, FAPERJ, FAPERGS, and FAPESP (Brazil); MES and BNSF (Bulgaria); CERN; CAS, MoST, and NSFC (China); MINCIENCIAS (Colombia); MSES and CSF (Croatia); RIF (Cyprus); SENESCYT (Ecuador); MoER, ERC PUT and ERDF (Estonia); Academy of Finland, MEC, and HIP (Finland); CEA and CNRS/IN2P3 (France); BMBF, DFG, and HGF (Germany); GSRI (Greece); NKFIH (Hungary); DAE and DST (India); IPM (Iran); SFI (Ireland); INFN (Italy); MSIP and NRF (Republic of Korea); MES (Latvia); LAS (Lithuania); MOE and UM (Malaysia); BUAP, CINVESTAV, CONACYT, LNS, SEP, and UASLP-FAI (Mexico); MOS (Montenegro); MBIE (New Zealand); PAEC (Pakistan); MES and NSC (Poland); FCT (Portugal); MESTD (Serbia); MCIN/AEI and PCTI (Spain); MOSTR (Sri Lanka); Swiss Funding Agencies (Switzerland); MST (Taipei); MHESI and NSTDA (Thailand); TUBITAK and TENMAK (Turkey); NASU (Ukraine); STFC (United Kingdom); DOE and NSF (USA).
\end{acknowledgments}

\clearpage

\bibliography{auto_generated} 

\cleardoublepage \appendix\section{The CMS Collaboration \label{app:collab}}\begin{sloppypar}\hyphenpenalty=5000\widowpenalty=500\clubpenalty=5000
\cmsinstitute{Yerevan Physics Institute, Yerevan, Armenia}
{\tolerance=6000
A.~Hayrapetyan, A.~Tumasyan\cmsAuthorMark{1}\cmsorcid{0009-0000-0684-6742}
\par}
\cmsinstitute{Institut f\"{u}r Hochenergiephysik, Vienna, Austria}
{\tolerance=6000
W.~Adam\cmsorcid{0000-0001-9099-4341}, J.W.~Andrejkovic, T.~Bergauer\cmsorcid{0000-0002-5786-0293}, S.~Chatterjee\cmsorcid{0000-0003-2660-0349}, K.~Damanakis\cmsorcid{0000-0001-5389-2872}, M.~Dragicevic\cmsorcid{0000-0003-1967-6783}, A.~Escalante~Del~Valle\cmsorcid{0000-0002-9702-6359}, P.S.~Hussain\cmsorcid{0000-0002-4825-5278}, M.~Jeitler\cmsAuthorMark{2}\cmsorcid{0000-0002-5141-9560}, N.~Krammer\cmsorcid{0000-0002-0548-0985}, D.~Liko\cmsorcid{0000-0002-3380-473X}, I.~Mikulec\cmsorcid{0000-0003-0385-2746}, J.~Schieck\cmsAuthorMark{2}\cmsorcid{0000-0002-1058-8093}, R.~Sch\"{o}fbeck\cmsorcid{0000-0002-2332-8784}, D.~Schwarz\cmsorcid{0000-0002-3821-7331}, M.~Sonawane\cmsorcid{0000-0003-0510-7010}, S.~Templ\cmsorcid{0000-0003-3137-5692}, W.~Waltenberger\cmsorcid{0000-0002-6215-7228}, C.-E.~Wulz\cmsAuthorMark{2}\cmsorcid{0000-0001-9226-5812}
% ...
\par}
\vskip\cmsinstskip
\dag:~Deceased\\
$^{1}$Also at Yerevan State University, Yerevan, Armenia\\
$^{2}$Also at TU Wien, Vienna, Austria\\
% ...
\end{sloppypar}
\end{document}